\documentclass[twocolumn,showpacs,aps,prd]{revtex4-1}
\usepackage{graphicx}
\usepackage{float}
\usepackage{epsfig,graphics,subfigure,psfrag,amsmath,amssymb}
\usepackage{lineno}
\usepackage{dcolumn}
\usepackage{bm}
\usepackage{overpic}
\usepackage{xspace}
\usepackage{rotating}
\usepackage{epstopdf}
\usepackage[colorlinks,linkcolor=blue,anchorcolor=blue,citecolor=blue]{hyperref}

\begin{document}
\title{\bf \boldmath Search for the rare decay $\eta'\rightarrow\pi^{0}\pi^{0}\pi^{0}\pi^{0}$ at BESIII}

\author{
  \begin{small}
    \begin{center}
 M.~Ablikim$^{1}$, M.~N.~Achasov$^{10,d}$, P.~Adlarson$^{59}$, S. ~Ahmed$^{15}$, M.~Albrecht$^{4}$, M.~Alekseev$^{58A,58C}$, A.~Amoroso$^{58A,58C}$, F.~F.~An$^{1}$, Q.~An$^{55,43}$, Y.~Bai$^{42}$
% ...
      \\
      \vspace{0.2cm}
      (BESIII Collaboration)\\
      \vspace{0.2cm} {\it
$^{1}$ Institute of High Energy Physics, Beijing 100049, People's Republic of China\\
% ...
$^{4}$ Bochum Ruhr-University, D-44780 Bochum, Germany\\
% ...
$^{10}$ G.I. Budker Institute of Nuclear Physics SB RAS (BINP), Novosibirsk 630090, Russia\\
% ...
$^{15}$ Helmholtz Institute Mainz, Johann-Joachim-Becher-Weg 45, D-55099 Mainz, Germany\\
% ...
$^{42}$ Southeast University, Nanjing 211100, People's Republic of China\\
$^{43}$ State Key Laboratory of Particle Detection and Electronics, Beijing 100049, Hefei 230026, People's Republic of China\\
% ...
$^{55}$ University of Science and Technology of China, Hefei 230026, People's Republic of China\\
% ...
$^{58}$ (A)University of Turin, I-10125, Turin, Italy; (B)University of Eastern Piedmont, I-15121, Alessandria, Italy; (C)INFN, I-10125, Turin, Italy\\
$^{59}$ Uppsala University, Box 516, SE-75120 Uppsala, Sweden\\
% ...
$^{d}$ Also at the Novosibirsk State University, Novosibirsk, 630090, Russia\\
% ...
      }
    \end{center}
    \vspace{0.4cm}
  \end{small}
}
\noaffiliation{}

\begin{abstract}
Using a sample of $1.31 \times 10^{9}$ $J/\psi$ events collected with the BESIII detector, we perform a search for the rare decay $\eta'\rightarrow 4\pi^{0}$ via $J/\psi\rightarrow\gamma\eta'$. No significant $\eta'$ signal is observed in the 4$\pi^{0}$ invariant mass spectrum. With a Bayesian approach, the upper limit on the branching fraction is determined to be $\mathcal{B}(\eta'\rightarrow 4\pi^{0})$ $< 4.94\times10^{-5}$ at the 90\% confidence level, which is a factor of six smaller than the previous experimental limit.
\end{abstract}

\maketitle
\section{Introduction}
The $\eta^\prime$ meson has a special role in improving the understanding of  low-energy Quantum Chromodynamics (QCD),  and studies of its decays have attracted considerable theoretical and experimental attention~\cite{kang:etap, huijuan:etap}. In addition to its important role in testing the fundamental discrete symmetries and searching for processes beyond the Standard Model (SM), $\eta^\prime$ decays offer unique opportunities to test chiral perturbation theory (ChPT) and the vector-meson dominance (VMD) model.

\begin{figure}[htb]
  \centering
  \includegraphics[width=0.5\textwidth]{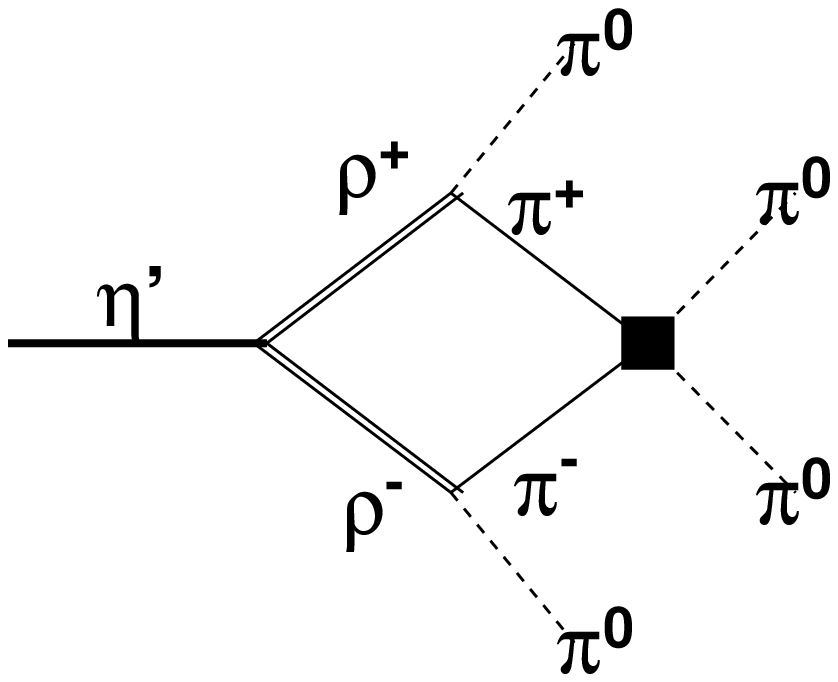}\put (-130,5){(a)}\\
  \includegraphics[width=0.5\textwidth]{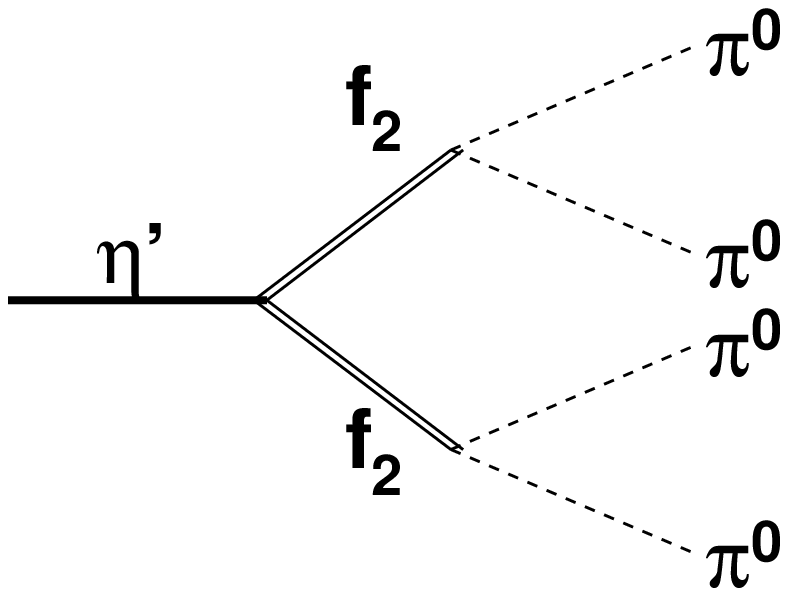}\put (-130,15){(b)}\\
  \caption{ D-wave pion-loop  (a) and intermediate $f_{2}$ mesons contribution (b) to $\eta^\prime\rightarrow4\pi^{0}$~\cite{theory4pi0}.
   }
  \label{fig:feyman}
\end{figure}

In theory, $\eta^\prime\rightarrow 4\pi^{0}$ is a highly suppressed decay because of the S-wave \emph{CP}-violation. In the light of an effective chiral Lagrangian approach, the S-wave \emph{CP}-violation in $\eta^\prime\rightarrow 4\pi^0$ is induced by the so-called $\theta$-term, which is an additional term in the QCD Lagrangian to account for the solution of the strong-\emph{CP} problem. It was found that the S-wave \emph{CP}-violation effect contributed to this decay is at a level of $10^{-23}$ \cite{br-CPviolating, theta}, which is far beyond current experimental sensitivity. However, higher-order contributions, involving a D-wave pion loop or the production of two $f_2$ tensor mesons (see Fig.~\ref{fig:feyman}), provide a \emph{CP}-conserving route through which the decay can occur. By ignoring the tiny contribution from the latter process, calculations based on ChPT and VMD models predict the branching fraction caused by D-wave \emph{CP}-conserving to be at the level of $10^{-8}$~\cite{theory4pi0}. Therefore, an observation of $\eta'\rightarrow 4\pi^{0}$ with a branching fraction at a level of $10^{-8}$ would indicate ChPT and VMD models are reliable in calculating the decay $\eta^\prime(\eta)\rightarrow 4\pi^{0}$.

So far the decay $\eta^\prime\rightarrow 4\pi^{0}$ has not been observed. About three decades ago the first attempt to search for this mode was performed by the joint CERN-IHEP experiment and the upper limit on the branching fraction was determined to be $\mathcal{B}(\eta^\prime\rightarrow4\pi^{0})< 5 \times 10^{-4}$ at the 90\% confidence level (C.L.)~\cite{CERN-IHEP}. The more recent upper limit of $3.2 \times 10^{-4}$ at 90\% C.L. was obtained by the  GAMS-4$\pi$ experiment~\cite{gams4pi}.

Although $\eta^\prime$ mesons can not be produced directly in $e^+e^-$ annihilations, the decay $J/\psi \rightarrow\gamma\eta^\prime$, with a branching fraction of $(5.13\pm0.17)\times 10^{-3}$~\cite{pdg2018}, provides an abundant source of $\eta^\prime$ meson in this environment.  The BESIII experiment has exploited this production mode to perform a series of studies of $\eta^\prime$ decays~\cite{ssfang},  based on a sample of $(1310.6\pm7.0)\times10^{6}~J/\psi$ events~\cite{jpsinumber}, corresponding to $6.7\times 10^6$ $\eta^\prime$ events.  In this paper, using this same $J/\psi$ sample, we perform a search for $\eta^\prime\rightarrow 4\pi^{0}$ via $J/\psi\rightarrow\gamma\eta^\prime$.

\section{BESIII DETECTOR AND MONTE CARLO SIMULATION}
The BESIII detector~\cite{Ablikim:2009aa} is a magnetic spectrometer located at the Beijing Electron Positron Collider (BEPCII)~\cite{Yu:IPAC2016-TUYA01}. The cylindrical core of the BESIII detector consists of a helium-based multilayer drift chamber (MDC), a
plastic scintillator time-of-flight system (TOF), and a CsI(Tl) electromagnetic calorimeter (EMC), which are all enclosed in a superconducting solenoidal magnet providing a 1.0~T (0.9~T in 2012) magnetic field. The solenoid is supported by an octagonal flux-return yoke with resistive plate counter muon-identifier modules interleaved with steel. The acceptance of charged particles and photons is 93\% over the $4\pi$ solid angle. The charged-particle momentum resolution at $1~{\rm GeV}/c$ is $0.5\%$, and the $dE/dx$ resolution is $6\%$ for electrons from Bhabha scattering. The EMC measures photon energies with a resolution of $2.5\%$ ($5\%$) at $1$~GeV in the barrel (end cap) region. The time resolution of the TOF barrel part is 68~ps, while that of the end cap part is 110~ps.

Simulated samples produced with a {\sc geant4}-based~\cite{geant4} Monte Carlo (MC) simulation framework, 
which includes the geometric description of the BESIII detector and the detector response, are used to determine the detection efficiency and to estimate the backgrounds. The simulation includes the beam energy spread and initial state radiation (ISR) in $e^+e^-$ annihilation modeled with the generator {\sc kkmc}~\cite{ref:kkmc}. The inclusive MC sample consists of the production of the $J/\psi$ resonance, and the continuum processes incorporated in {\sc kkmc}~\cite{ref:kkmc}. The known decay modes are modeled with {\sc evtgen}~\cite{ref:evtgen} using branching fractions taken from the
Particle Data Group (PDG)~\cite{pdg2018}, and the remaining unknown decays of the charmonium states with {\sc lundcharm}~\cite{ref:lundcharm}. The final state radiation (FSR) from charged final-state particles are incorporated with the {\sc photos} package~\cite{photos}.

\section{EVENT SELECTION}

In this analysis, the pseudoscalar mesons $\eta^\prime$ and $\pi^0$ are reconstructed in the modes $\eta^\prime\rightarrow 4\pi^0$ and $\pi^0\rightarrow\gamma\gamma$. Candidate $J/\psi\rightarrow\gamma 4\pi^0$ decays are chosen by selecting events with at least nine isolated photons and no charged tracks. Photon candidates are reconstructed from clusters of energy deposited in the EMC. Photon candidates are required to have at least 25 MeV of energy for barrel showers ($| \cos \theta| \leq 0.8$) or 50 MeV for end-cap showers ($0.86 \leq|\cos \theta| \leq0.92$). 

\begin{figure}[htb]
  \centering
  \includegraphics[width=0.4\textwidth]{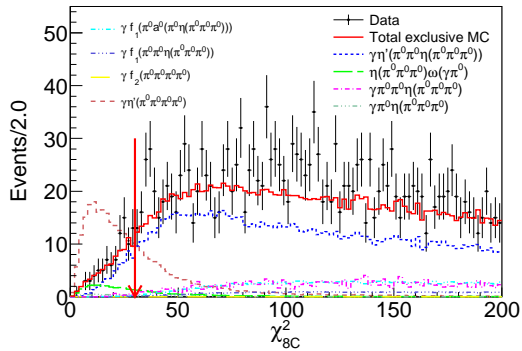}\\
  \caption{$\chi^{2}_{\rm 8C}$ distribution of candidate events in the $\eta^\prime$ signal region showing data and the contribution from the considered background sources. The arrow indicates the selection requirement of $\chi^{2}_{\rm 8C}<30$.}
  \label{fig:election}
\end{figure}

The photon candidate with the maximum energy deposited in the EMC is treated as the radiative photon directly originating from the $J/\psi$ decay. For the two-body signal decay $J/\psi\rightarrow\gamma\eta^\prime$, this photon carries a unique energy of 1.4 GeV.
To reconstruct the $\pi^0$ candidate from the remaining photons, a one-constraint (1C) kinematic fit is performed to each photon pair with the invariant mass constrained to the $\pi^{0}$ mass, with the requirement that the goodness-of-fit $\chi_{\rm 1C}^{2}(\gamma\gamma)<25$. Then an eight-constraint (8C) kinematic fit is performed for the $\gamma\pi^0\pi^0\pi^0\pi^0$ combination by enforcing energy-momentum conservation and constraining the invariant masses of each of the four photon pairs to the nominal $\pi^0$ mass. If more than one combination is found in an event, only the one with the smallest $\chi^{2}_{\rm 8C}$ is retained. The $\chi^{2}_{\rm 8C}$ distribution is shown in Fig.~\ref{fig:election}. Candidate events with $\chi^{2}_{\rm 8C}>30$ are rejected.

To ensure a good description of data, a signal MC simulation is modeled with the decay amplitude in Ref.~\cite{theory4pi0}, which assumes the pion-loop contribution as shown in Fig.~\ref{fig:feyman}(a)  (the contribution from $f_2$ mesons shown in  Fig.~\ref{fig:feyman}(b) is considered to be much smaller). After the full event selection, the detection efficiency ($\varepsilon$) is determined to be 2.46\%.

Figure~\ref{Fig:fitMC} shows the 4$\pi^0$ mass spectrum, $M(4\pi^0)$, after selection. No significant $\eta^\prime$ signal is evident.

In order to investigate possible sources of contamination we apply the selection to an inclusive MC sample of 1.2 $\times 10^{9}$ J/$\psi$ events.  All the significant background components are listed in  Table \ref{Tab:exMC}, showing their expected contribution to the data sample. Since the decay $J/\psi\rightarrow n\pi^0$, where $n$ is the number of $\pi^{0}$ mesons, is forbidden because of $C$ conservation, we find that the background comes mainly from decays with the same final state as the signal, for example $J/\psi\rightarrow\eta\omega, \,\eta\rightarrow\pi^{0}\pi^{0}\pi^{0}, \, \omega\rightarrow\gamma\pi^{0}$ or from radiative decays with more than four $\pi^0$s in the final states, of which the dominant mode is $J/\psi\rightarrow\gamma\eta^\prime, \, \eta^\prime\rightarrow\pi^{0}\pi^{0}\eta, \, \eta\rightarrow \pi^{0}\pi^{0}\pi^{0}$.  This latter channel contributes to a broad structure around 0.88~GeV/c$^{2}$ in  the 4$\pi^{0}$ mass spectrum as indicated by the dashed line in Fig.~\ref{Fig:fitMC}.   No other source of background peaks in the $\eta^\prime$ mass region.

\begin{table}[htb]
  \centering
  \caption{The main background channels and their expected contribution to the selected sample.}
  \vspace{0.05in}
  \label{Tab:exMC}
  \begin{tabular}{l p{2.6cm}<{\centering}}
    \hline
    \hline
    \small Decay mode&  \small Number of events \\
    \hline
    \small$J/\psi\rightarrow\gamma\eta^\prime, \, \eta^\prime\rightarrow\pi^{0}\pi^{0}\eta,\, \eta\rightarrow\pi^{0}\pi^{0}\pi^{0}$ & \small 496 \\
    \small$J/\psi\rightarrow\eta\omega,\, \eta\rightarrow\pi^{0}\pi^{0}\pi^{0},\, \omega\rightarrow\gamma\pi^{0}$ &  \small 131 \\
     \small $J/\psi\rightarrow\gamma \pi^{0}\pi^{0}\eta,\,\eta\rightarrow\pi^{0}\pi^{0}\pi^{0}$ &\small 38 \\
    \small $J/\psi\rightarrow\gamma \pi^{0}\eta, \,\eta\rightarrow\pi^{0}\pi^{0}\pi^{0}$ &\small 24 \\
    \small $J/\psi\rightarrow\gamma f_{1}(1285), \, f_{1}(1285)\rightarrow\pi^{0} a^{0}$& \small 11\\
    \small $J/\psi \rightarrow \gamma f_{2}(1270), \,f_{2}(1270)\rightarrow \pi^{0}\pi^{0}\pi^{0}\pi^{0}$ & \small 10 \\
    \small $J/\psi\rightarrow\gamma f_{1}(1285), \, f_{1}(1285)\rightarrow\pi^{0}\pi^{0}\eta$ & \small 5 \\
    \hline
    \hline
  \end{tabular}
  \end{table}

\section{FIT MODEL AND UPPER LIMIT ON BRANCHING FRACTION OF $\eta'\rightarrow 4\pi^{0}$ }

\begin{figure}[htb]
  \centering
  \includegraphics[width=0.4\textwidth]{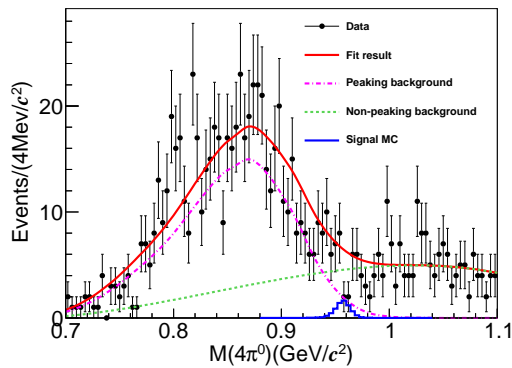}\\
    \caption{The M$(4\pi^{0})$ distribution in data, together with the total fit result and the contributions from non-peaking background and the peaking background $J/\psi\rightarrow\gamma\eta^\prime, \, \eta^\prime\rightarrow\pi^{0}\pi^{0}\eta,\, \eta\rightarrow\pi^{0}\pi^{0}\pi^{0}$. Also shown is the expected shape of the signal contribution, with arbitrary normalisation.}
  \label{Fig:fitMC}
\end{figure}

An unbinned maximum likelihood fit is performed on the $4\pi^0$ mass spectrum, allowing for background contributions and a possible signal component.  The peaking background  $J/\psi\rightarrow\gamma\eta^\prime, \, \eta'\rightarrow\pi^{0}\pi^{0}\eta,\, \eta\rightarrow\pi^{0}\pi^{0}\pi^{0}$ is described by a distribution obtained from a dedicated MC generator~\cite{generator_eta2pi0, generator_3pi0}, and its normalisation fixed from knowledge of the number of $J/\psi$ events and the branching fractions of the decays. The non-peaking background contribution is modeled with a third-order Chebychev polynomial function. The signal shape and resolution is taken from simulation. The fit result is shown superimposed in Fig.~\ref{Fig:fitMC}, where the signal contribution is negligible.

A Bayesian approach is used to determine an upper limit on the branching fraction of $\eta'\rightarrow 4\pi^{0}$. Many fits are performed for different assumed values of the signal yield $N$, which is a fixed parameter, and for each fit the negative log-likelihood ${\cal {S}}$ is determined. For each value of $N$ the branching fraction is
\begin{equation}\label{Branching}
 \mathcal{B}(\eta^\prime\rightarrow 4\pi^{0})=\frac{N}{N_{J/\psi} \cdot \varepsilon \cdot \mathcal{B}(J/\psi\rightarrow\gamma\eta^\prime)\cdot
 \mathcal{B}(\pi^{0}\rightarrow\gamma\gamma)^{4}},
\end{equation}
where $N_{J/\psi} = (1310.6\pm7.0)\times10^{6}$  is the number of $J/\psi$ events~\cite{jpsinumber},
$\varepsilon$ is the detection efficiency, $\mathcal{B}(J/\psi\rightarrow\gamma\eta^\prime)$ and $\mathcal{B}(\pi^{0}\rightarrow\gamma\gamma)$ are the branching fractions of $J/\psi\rightarrow\gamma\eta^\prime$ and $\pi^{0}\rightarrow\gamma\gamma$, respectively, which are taken from Ref.~\cite{pdg2018}.

The distribution of normalized likelihood values, defined as ${\cal L(B)} = \exp(-[{\cal S(B)} - {\cal S}_{\rm min}])$, where ${\cal S}_{\rm min}$ is the lowest negative log-likelihood obtained from the ensemble of fits,  is taken as the probability density function (PDF) for the expected branching fraction of $\eta^\prime\rightarrow 4\pi^{0}$. The upper limit on the branching fraction at the 90\% C.L., defined as $\mathcal{B}_{UL}$, corresponds to the branching fraction at 90\% of the integral of the PDF,
\begin{equation}\label{upperlimit}
  \frac{\int_{0}^{\mathcal{B}_{UL}}\mathcal{L}(\mathcal{B})d\mathcal{B}}{\int_{0}^{\infty}\mathcal{L}(\mathcal{B})d\mathcal{B}}=0.9,
\end{equation}
and is found to be $4.57 \times 10^{-5}$, considering statistical uncertainties alone.

\section{SYSTEMATIC UNCERTAINTIES}

Two categories of systematic uncertainty are considered: those associated with the fit model and procedure, and those which enter when using Eq.~(\ref{Branching}) to express the signal yield as a branching fraction.

The fit-related uncertainties come mainly from the fitting ranges, signal shape, non-peaking background shape, peaking background shape and the number of the peaking background events.

The systematic uncertainty from the fit ranges is estimated by varying them by $\pm5$ MeV/c$^2$.

In the fit to the M(4$\pi^{0}$) distribution, signal shape is taken from MC simulation. To assess the uncertainty due to the signal shape, an alternative fit is performed by convolving a Gaussian function with a fixed resolution of 2.6 MeV and mean of 2.1 MeV which are obtained from a high purity control sample of $J/\psi \rightarrow\gamma\eta^\prime, \eta^\prime\rightarrow\pi^{0}\pi^{0}\eta, \eta\rightarrow\gamma\gamma$.

The uncertainty from the non-peaking background shape is determined by using a fourth-order Chebychev polynomial in place of the third-order Chebychev polynomial.

To assess the uncertainty associated with the number of the peaking background events, its contribution is re-calculated after varying the branching fractions of $J/\psi \rightarrow\gamma\eta^\prime$ and its cascade decays, $\eta'\rightarrow\pi^{0}\pi^{0}\eta$ and $\eta\rightarrow \pi^{0}\pi^{0}\pi^{0}$, within their uncertainties, and new fits are performed.

The systematic uncertainty associated with peaking background shape is evaluated by convolving a Gaussian function with resolution and mean value left free.

Among these cases, the dominant fit-model uncertainty arises from fitting range [0.705, 1.095] GeV/c$^{2}$ and it changes the upper limit at the 90\% C.L. to $\mathcal{B}_{UL}=4.88\times 10^{-5}$.

The other category of systematic uncertainties, summarised in Table~\ref{tab:systematic}, has contributions from the knowledge the photon detection efficiency, the efficiency of the kinematic fit, signal model, the branching fractions of the sub-decays involved in the signal process, and the total number of $J/\psi$ events.

The uncertainty from the photon detection is investigated with a high purity control sample of $J/\psi\rightarrow\pi^{+}\pi^{-}\pi^{0}$.
It is found that the differences between data and MC simulation are 0.5\% and 1.5\% for each photon deposited in the barrel and end cap of the EMC, respectively. With the same approach as used in Ref. \cite{systematic1}, the uncertainty on the detection efficiency for each photon in the signal decay is estimated to be 0.53\%, and thus the nine photons in the final state induce an overall uncertainty of 4.8\%.

The uncertainty associated with the kinematic fit is estimated by adjusting the components of the photon-energy error matrix in the signal MC sample to reflect the known difference in resolution between data and MC simulation~\cite{systematic3}. From the study of $\psi(3686)\rightarrow\gamma\chi_{c1} (\chi_{c1}\rightarrow 4\pi^0)$ decay\cite{systematic2}, it is known that the energy resolution in data is 4\% wider than in MC simulation. The relative difference in efficiency, 4.1\%, is taken as the systematic uncertainty from the kinematic fit.

In the normal fit, the cascade decay $\eta^\prime\rightarrow4\pi^{0}$ is described with the decay amplitude in Ref.~\cite{theory4pi0}. A fit with an alternative signal model replacing $\eta^\prime\rightarrow4\pi^{0}$ decay amplitude with a phase space (PHSP) distribution is performed. The change of the efficiency, 2.0\%, is taken as the uncertainty due to the signal model.

The relative uncertainty in the knowledge of  the branching fractions of  $J/\psi\rightarrow\gamma\eta^\prime$ and $\pi^{0}\rightarrow\gamma\gamma$~\cite{pdg2018} induces a corresponding uncertainty on the calculated upper limit on the branching fraction. The number of $J/\psi$ events is determined from the measured number of hadronic decays, and is found to be $(1310.6\pm7.0)\times10^{6}$~\cite{jpsinumber}, which corresponds to a relative uncertainty of 0.54\%.

Assuming all systematic uncertainties presented in Table \ref{tab:systematic} are independent, the total relative uncertainty is obtained to be 7.3\%, by adding all individual uncertainties in quadrature.

\begin{table}[htb]
  \centering
  \caption{Summary of the systematic uncertainties unrelated to the fit model.
For each component the relative impact on the branching fraction is listed in \%.}
  \vspace{0.1in}
  \renewcommand\arraystretch{1.0}
   \begin{tabular}{p{5.0cm}p{3.5cm}<{\centering}}
     \hline\hline
     Source  & Systematic uncertainties\\
     \hline
     Photon detection & 4.8 \\
     Kinematic fit  &  4.1 \\
     Signal model   &  2.0 \\
     $\mathcal{B}(J/\psi\rightarrow\gamma\eta^\prime)$ &  3.1 \\
      $\mathcal{B}(\pi^0\rightarrow\gamma\gamma)$ &  0.03 \\
     Number of $J/\psi$ events &  0.54 \\
     Total &  7.3 \\
     \hline\hline
   \end{tabular}
  \label{tab:systematic}
\end{table}

\section{RESULT}

The final upper limit on the branching fraction is determined by convolving the likelihood distribution $\mathcal{L}$ with the systematic uncertainties to obtain the smeared likelihood $\mathcal{L}^{\rm smear}$.
\begin{equation}
\label{Eq2}
 \mathcal{L}^{\rm smear}(\mathcal{B}) = \int \mathcal{L}\left(\frac{\varepsilon}{\overline{\varepsilon}}\mathcal{B}\right)\exp\left(-\frac{(\varepsilon - \overline{\varepsilon})^{2}}{2\sigma^{2}_{\varepsilon}}\right) d\varepsilon.
\end{equation}
In this exercise all components listed in Table~\ref{tab:systematic}, whatever their nature, can be considered as an uncertainty on the detection efficiency $\varepsilon$. The nominal efficiency value is $\overline{\varepsilon}$, $\sigma_{\varepsilon}$ is the absolute total systematic uncertainty on the efficiency, and $\mathcal{B}$ is the branching fraction of $\eta^\prime\rightarrow 4\pi^{0}$.

\begin{figure}[htb]
  \centering
  \includegraphics[width=0.4\textwidth]{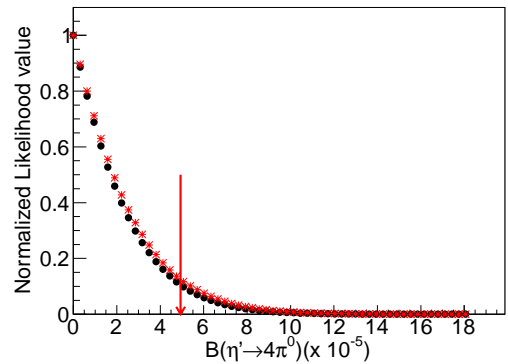}\\
  \caption{Normalized likelihood distribution before (black dots) and after (red stars) convolution with systematic uncertainty.}
  \label{likelihood2}
\end{figure}

Figure \ref{likelihood2} shows the normalized likelihood distribution after taking all systematic uncertainties into account. 
The corresponding upper limit of the branching fraction of $\eta^\prime\rightarrow4\pi^{0}$ at the 90\% C.L. is determined to be $4.94\times10^{-5}$.

\section{Summary}

Using a sample of $1.31 \times 10^{9} J/\psi$ events collected with the BESIII detector, a search for the decay $\eta^\prime\rightarrow4\pi^{0}$ is performed via $J/\psi\rightarrow\gamma\eta'$. No evidence for the rare decay $\eta^\prime\rightarrow4\pi^0$ is found, and an upper limit of $\mathcal{B}(\eta^\prime\rightarrow 4\pi^0)<4.94\times 10^{-5}$ is set at the 90\% confidence level. This limit is approximately a factor of six smaller than the previous most stringent result~\cite{gams4pi}.

The current limit is still far above the level of $10^{-8}$, which is predicted in theory. Further studies of $\eta^\prime$ rare decays are still necessary to test ChPT and VMD model and look for the \emph{CP}-violation (S-wave) $\eta\rightarrow 4\pi^{0}$ decay. A sample of  $10^{10}$ $J/\psi$ events has now been collected at BESIII, which will allow for even more sensitive searches to be performed for this important decay mode.

\section{ACKNOWLEDGMENTS}
The BESIII collaboration thanks the staff of BEPCII and the IHEP computing center for their strong support.
This work is supported in part by National Key Basic Research Program of China under Contract No. 2015CB856700;
National Natural Science Foundation of China (NSFC) under Contracts Nos. 11335008, 11425524, 11625523, 11635010, 11675184, 11735014;
the Chinese Academy of Sciences (CAS) Large-Scale Scientific Facility Program;
the CAS Center for Excellence in Particle Physics (CCEPP);
Joint Large-Scale Scientific Facility Funds of the NSFC and CAS under Contracts Nos. U1532257, U1532258, U1732263;
CAS Key Research Program of Frontier Sciences under Contracts Nos. QYZDJ-SSW-SLH003, QYZDJ-SSW-SLH040;
100 Talents Program of CAS;
INPAC and Shanghai Key Laboratory for Particle Physics and Cosmology;
German Research Foundation DFG under Contracts Nos. Collaborative Research Center CRC 1044, FOR 2359;
Istituto Nazionale di Fisica Nucleare, Italy;
Koninklijke Nederlandse Akademie van Wetenschappen (KNAW) under Contract No. 530-4CDP03;
Ministry of Development of Turkey under Contract No. DPT2006K-120470;
National Science and Technology fund;
The Swedish Research Council;
U. S. Department of Energy under Contracts Nos. DE-FG02-05ER41374, DE-SC-0010118, DE-SC-0010504, DE-SC-0012069;
University of Groningen (RuG) and the Helmholtzzentrum f{\"u}r Schwerionenforschung GmbH (GSI), Darmstadt.

\end{document}